\documentstyle[12pt]{article}
 \makeatletter

\def\@authoraddress{}
\def\@title{}
\def\title#1{\gdef\@title{{\par\vskip-10pt\Large\bf
\baselineskip20pt\centering\ignorespaces\uppercase{#1}\vskip6pt}}%
\setcounter{table}{0}      \setcounter{figure}{0}
\setcounter{equation}{0}   \setcounter{section}{0}
\setcounter{subsection}{0} \setcounter{subsubsection}{0}
\setcounter{paragraph}{0}
}

\def\authors#1{\expandafter\def\expandafter\@authoraddress\expandafter
{\@authoraddress %
{\dimen0=-\prevdepth \advance\dimen0 by1.5\baselineskip
\nointerlineskip \centering
\vrule height\dimen0 width0pt\relax\ignorespaces\large\sc#1\par
}%
}%
}

\def\addresses#1{\expandafter\def\expandafter\@authoraddress\expandafter
{\@authoraddress{\nointerlineskip\vskip1pc
                 \footnotesize\it\centering\ignorespaces#1\par}}}

\def\@maketitle{%
\@title
\ifdim\prevdepth=-1000pt \prevdepth0pt\fi
\@authoraddress
}

\def\maketitle{\par
\begingroup
\let\cite\@bylinecite
\global\@topnum\z@ %
\@maketitle
\endgroup
\def\@thanks{}\def\@authoraddress{}\def\@title{}
}

\def\abstract{\par
\bgroup
\ifdim\prevdepth=-1000pt \prevdepth0pt\fi
\hsize\columnwidth
\leftskip=2em \rightskip\leftskip
\dimen0=-\prevdepth \advance\dimen0 by2pc \nointerlineskip
\noindent\vskip1.5\baselineskip\nointerlineskip\noindent\footnotesize\relax}

\newif\if@firststuff

\def\endabstract{\par
\nointerlineskip \vskip0pt
\noindent \par
\egroup
\hrule depth0pt width0pt
\global\everypar{\global\@firststufffalse}\global\@firststufftrue
}

\renewcommand\section{\@startsection {section}{1}{\z@}%
                                   {-3.5ex \@plus -1ex \@minus -.2ex}%
                                   {2.3ex \@plus.2ex}%
                                   {\normalfont\large\bfseries}}
\renewcommand\subsection{\@startsection{subsection}{2}{\z@}%
                                     {-3.25ex\@plus -1ex \@minus -.2ex}%
                                     {1.5ex \@plus .2ex}%
                                     {\normalfont\large\bfseries}}

\def\1ad{\mbox{\normalsize $^1$}}
\def\2ad{\mbox{\normalsize $^2$}}
\def\3ad{\mbox{\normalsize $^3$}}
\def\4ad{\mbox{\normalsize $^4$}}
\def\5ad{\mbox{\normalsize $^5$}}
\def\6ad{\mbox{\normalsize $^6$}}
\def\7ad{\mbox{\normalsize $^7$}}
\def\8ad{\mbox{\normalsize $^8$}}
\def\adref#1{\mbox{\normalsize $^{#1}$}}
\makeatother

\renewcommand{\r}{\rho} %!!!!!!!!!!!!!!
\advance\voffset by 15mm   %POUR CBPF
\makeatletter
\@addtoreset{equation}{section}
\makeatother

\def\ftoday{{\sl {Le \number\day \space\ifcase\month 
\or janvier\or f\'evrier\or mars\or avril\or mai
\or juin\or juillet\or ao\^ut\or septembre\or octobre
\or novembre \or d\'ecembre\fi\space \number\year}}}    
\def\ptoday{{\sl {\number\day \space de\space \ifcase\month 
\or janeiro\or fevereiro\or mar{\c c}o\or abril\or maio
\or junho\or julho\or agosto\or setembro\or outubro
\or novembro \or dezembro\fi\space de\space \number\year}}}    
\def\gtoday{{\sl {Den \number\day. \ifcase\month 
\or Januar\or Februar\or M\"arz\or April\or Mai
\or Juni\or Juli\or August\or September\or Oktober
\or November \or Dezember\fi\space \number\year}}}    
\def\today{{\sl {\ifcase\month
\or January\or February\or March\or April\or May
\or June\or July\or August\or September\or October
\or November \or December\fi \space\number\day,\space 
                                            \number\year}}}
\newcommand{\journal}[4]{{\em #1~}#2\,(#3)\,#4}

\newcommand{\ijmp}{\journal {Int. J. Mod. Phys.}}

\newcommand{\pr}{\journal {Phys. Rev.}}

\newcommand{\jmp}{\journal {J. Math. Phys.}}

\newcommand{\cmp}{\journal {Commun. Math. Phys.}}

\newcommand{\cqg}{\journal {Class. Quantum Grav.}}

\newcommand{\np}{\journal {Nucl. Phys.}}
\newcommand{\pl}{\journal {Phys. Lett.}}

\newcommand{\prep}{\journal {Phys. Rep.}}

\newcommand{\nc}{\journal {Nuovo Cimento}}

\renewcommand{\d}{\delta}         
\newcommand{\e}{\varepsilon}

\newcommand{\la}{\lambda}        
\newcommand{\m}{\mu}
\newcommand{\n}{\nu}
\newcommand{\om}{\omega}

           \newcommand{\F}{{\Phi}}
\newcommand{\vf}{{\varphi}}
\renewcommand{\AA}{{\cal A}}

\newcommand{\LL}{{\cal L}}

\newcommand{\es}{\\[3mm]}

\newcommand{\na}{\nabla}

\newcommand{\sla}{\raise.15ex\hbox{$/$}\kern -.57em} 
\newcommand{\Sla}{\raise.15ex\hbox{$/$}\kern -.70em}

\newcommand{\lp}{\left(}\newcommand{\rp}{\right)}
\newcommand{\lc}{\left[}\newcommand{\rc}{\right]}

\newcommand{\complex}{{\kern .1em {\raise .47ex
\hbox {$\scriptscriptstyle |$}}
    \kern -.4em {\rm C}}}
\newcommand{\real}{{{\rm I} \kern -.19em {\rm R}}}
\newcommand{\rational}{{\kern .1em {\raise .47ex
\hbox{$\scripscriptstyle |$}}
    \kern -.35em {\rm Q}}}
\renewcommand{\natural}{{\vrule height 1.6ex width
.05em depth 0ex \kern -.35em {\rm N}}}

\newcommand{\tr}{{\rm {Tr} \,}}
\newcommand{\half}{\dfrac{1}{2}}
\newcommand{\pa}{\partial}

\newcommand{\dfud}[2]{{\displaystyle{\frac{\delta #1}{\delta #2}}}}
\newcommand{\dfrac}[2]{{\displaystyle{\frac{#1}{#2}}}}
\newcommand{\dsum}[2]{\displaystyle{\sum_{#1}^{#2}}}   
\newcommand{\dint}{\displaystyle{\int}}

\newcommand{\twiddle}{\lower.9ex\rlap{$\kern -.1em\scriptstyle\sim$}}

\newcommand{\equ}[1]{(\ref{#1})}
\newcommand{\eq}{\begin{equation}}
\newcommand{\eqn}[1]{\label{#1}\end{equation}}
\newcommand{\eea}{\end{eqnarray}}
\newcommand{\eqa}{\begin{eqnarray}}
\newcommand{\eqan}[1]{\label{#1}\end{eqnarray}}
\newcommand{\ba}{\begin{array}}
\newcommand{\ea}{\end{array}}
\newcommand{\eqac}{\begin{equation}\begin{array}{rcl}}
\newcommand{\eqacn}[1]{\end{array}\label{#1}\end{equation}}

\newcommand{\nablag}{\nabla^{({\rm g})}}
\newcommand{\dse}{\d^{\,{\rm S}}_{(\e)}}
\newcommand{\sqg}{\sqrt{-g}}
\newcommand{\BC}{{\bar C}}
\newcommand{\bxi}{{\bar\xi}}
\newcommand{\bom}{{\bar\om}}
\newcommand{\dintm}{\displaystyle{\int_M}}
\begin{document}
{\hfill UFES-DF-OP2000/3  ---  hep-th/0010053}
\vspace{10mm}

\title{Diffeomorphism Invariant Theories and Vector Supersymmetry}

\authors{Olivier Piguet\footnote{Work supported in part by the 
{\it Conselho Nacional de Desenvolvimento Cient\'\i fico e 
Tecnol\'ogico (CNPq -- Brazil).}
\\ \indent Talk given at the
International Conference ``Quantization, Gauge Theory, and Strings''
dedicated to the memory of Professor Efim Fradkin, Moscow, 2000.}
\adref{}}

\addresses{Universidade Federal do Esp\'{\i}rito Santo (UFES)
, \\CCE, Departamento de F\'{\i}sica,\\ Campus Universit\'ario
de Goiabeiras - BR-29060-900 - Vit\'oria - ES - Brasil.\\
{\tt E-mail: piguet@cce.ufes.br}}

\maketitle
%%%%%%%%%%%%%%%%%%%%%%%%%%%%%%%%%%%%%%%%%%%%%%%%%%%%%%%%%%%%%%%

\begin{abstract}
Einstein gravity in the Palatini first order formalism is
shown to possess a vector supersymmetry of the type encountered
in the topological gauge theories. A peculiar feature of
the gravitationel theory is the
link of this vector supersymmetry with the field equation of motion
of the Faddeev-Popov ghost associated to  
diffeomorphism invariance. 
\end{abstract}

%%%%%%%%%%%%%%%%%%%%%%%%%%%%%%%%%
\subsection*{1\ Introduction and Conclusions}

I think it is relevant, in a conference organized 
to the memory of Efim Fradkin, 
who devoted an important part of his work to the understanding of fundamental
symmetries, to talk about some not broadly known symmetry features 
of theories such as topological and gravitational theories. These
theories are well known to be characterized by their invariance 
under the diffeomorphims of the space-time manifold
(``general coordinate invariance''). Also, both types of theories
are gauge theories, the gauge group of gravity being the Lorentz
group. On the other hand, 
an interesting feature of topological theories such as 
Chern-Simons or $BF$ theory, is the presence of a
``vector supersymmetry'' -- a supersymmetry whose generator is a vector
valued operator~\cite{vector-susy}. 
In case the manifold admits isometries generated by 
Killing vectors -- e.g. the space-time translations 
if the background metric is 
flat -- the vector supersymmetry is a symmetry of the gauge-fixed
action. It happens that its generator together with the
BRST symmetry generator form an algebra which closes on the 
generators of the isometries of the translation type~\cite{cs-lucch-pig}. 
Vector supersymmetry
has been shown to play a key role in the ultraviolet finiteness of the
topological 
theories~\cite{UV-finiteness-vector-SUSY,alg-ren}.

The question addressed in the present talk is about the possibility of
such a vector supersymmetry in gravitation theory. 

In the Palatini or first order formalism of gravitation 
theory~\cite{peldan}, 
the independent dynamical variables 
are the vierbein 1-form $E$ (giving the metric $G$) 
and the connection 1-form $A$. 
The vierbein
dependence of the connection is given by its field equation, 
whereas Einstein equation results from the field equation with
respect to $E$. The action can be written, 
following~\cite{baez-knots,pullin-piren}, 
in a ``topological'' form, i.e. in such a way that
it can be interpreted as an action of the 1-form fields $E$ and $A$ on a
differentiable manifold $M$, without reference to any a-priori background
metric. The latter point is known~\cite{witten} to be an essential 
characteristic of  
topological theories, and trying to exploit this feature belongs to
 the spirit of the modern attempts 
towards a construction of quantum gravity 
(see~\cite{gaul-rovelli,pullin-piren} for reviews and further references). 

Both local symmetries -- diffeomorphism and local
Lorentz invariances -- have to be gauge fixed. 
We shall choose a gauge fixing of the Landau type, within
the BRST formalism~\cite{alg-ren}. Much in the same way 
as in topological theories, this requires the introduction of a
nondynamical, background metric $g$. It should be clear that the
background metric, being introduced only in the gauge fixing part of the
theory, should not affect in any way the physical 
outcome, as it has been explicitely shown in~\cite{cs-lucch-pig} 
for the  -- perturbative -- quantum version of the
Chern-Simons theory.

%%%%%%%%%%%%%%%%%%%%%%%%%%%%%%%%%%%%%%%%%%%%%%%%%%

An interesting feature of gauge theories in a Landau type gauge
is the so-called 
ghost equation~\cite{ghost-eq,alg-ren}, 
which restricts the coupling of the
ghosts and implies the nonrenormalization of their
field amplitude\footnote{A review 
of the properties of topological theories mentionned above may
be found in Chapters 6 and 7 of ref.~\cite{alg-ren}.}.
We shall see that the vector supersymmetry in the gravitational case
is a direct consequence of
the field equation of the Faddeev-Popov ghost associated to
diffeomorphism invariance.  

Vector supersymmetry will turn out to be maximal if the background
metric is flat~\cite{cs-lucch-pig}. 
Whereas, in this case, the supersymmetry generators 
of the topological theories are 
the components of a vector, the superalgebra closing on the translations,
the supersymmetry generators of Einstein gravity in the Palatini 
formalism will be seen to be the components of 
one vector and one antisymmetric tensor: the full algebra 
here contains the ten Poincar\'e generators.

The present work is concerned only with the 
classical aspects of the theory. However, the results are of interest
since they reveal the link between the construction 
of the observables via the $\delta$ operator of 
Sorella~\cite{sorella-delta} 
associated to vector symmetry, on one hand, 
and diffeomorphism invariance and the corresponding ghost equation,
on the other hand.

A more detailed account of the results presented here
may be found in~\cite{piguet-cqg}.

%%%%%%%%%%%%%%%%%%%%%%%%%%%%%%%%%
\subsection*{2\ Action for Gravity in the Palatini Formalism}

The Einstein gravity Lagrangian in the first order formalism of 
Palatini~\cite{peldan} may be written as~\cite{baez-knots}:
\eq
S_{\rm inv} =
 \dfrac{1}{4}\dintm 
\e_{IJKL} E^I\wedge E^J \wedge F^{KL}(A) + S_{\rm matter}(E,A,\F)\ .
\eqn{inv-action}
The integral is taken over some differentiable manifold $M$, 
$E^I$ is a vierbein 1-form, with $I=0,\cdots,3$ a tangent 
plane Lorentz index. $F^{KL}$ is the curvature 2-form
\eq
F^{IJ}(A) = d A^{IJ} + A^{JK}\wedge A_K{}^J
\eqn{curv} 
of a connection\footnote{If the connection is self-dual, \equ{inv-action}
is the Ashtekar action~\cite{ashtekar,baez-knots}.}
$A^{IJ}$, the latter being taken as an independant  
variable\footnote{In a particular coordinate frame with 
$x =$ $(x^\m,\,\m=0,\cdots,3)$, $E^I = E_\m^I dx^\m$,
$F^{IJ} = \half F_{\m\n}^{IJ} dx^\m\wedge dx^\n$, etc.}. 
$\e_{IJKL}$ is the rank four antisymmetric tensor, normalized by
$\e_{0123}=1$. In the following, the  exterior multiplication
symbol $\wedge$ will be omitted. 
$S_{\rm matter}$ is some action for minimally coupled matter fields $\F$,
which we don't need to specify. We shall in fact omit this 
part in the following, for the sake of simplicity and without loss of
generality.

The field equations given by the variations of this action read
\eq\ba{l}
\dfud{S_{\rm inv}}{E^I} = \half\e_{IJKL} E^J F^{KL} \ , \es
\dfud{S_{\rm inv}}{A^{IJ}} = e_{IJKL} E^K DE^L \ ,
\ea\eqn{eq-motion}
where $D$ is the covariant exterior derivative:
$DE^I = dE^I - A^I{}_J E^J$.
It is known~\cite{peldan,baez-knots} 
that they lead to the usual specification of the connection as
a function of the vierbein and to the Einstein equation, in a Riemanian
space-time with metric
\eq
G_{\m\n}= E_\m^I E_\n^J \eta_{IJ}\ ,
\eqn{metric}
where $\eta_{IJ}$ is the flat metric used to 
lower and rise the tangent space indices $I,J,\cdots$.
 
The action \equ{inv-action} is invariant under the diffeomorphisms.
In infinitesimal form, they read
\eq
\d_{(\xi)} \vf = \LL_\xi \vf\ ,\quad \vf = E^I,\,A^{IJ}\ ,
\eqn{diff}
where the infinitesimal parameter is a vector field $\xi$ and
$\LL_\xi$ is the Lie derivative along $\xi$. 
The action is also invariant under 
the local Lorentz transformations which, in infinitesimal form, read
\eq\ba{l}
\d_{(\om)} E^I = \om^I{}_J E^J\ ,\es
\d_{(\om)} A^{IJ} =  d\om^{IJ} + \om^I{}_K A^{KJ} + \om^J{}_K A^{IK}\ ,
\ea\eqn{loc-lor}
with local parameters $\om^{IJ}=-\om^{JI}$.
%%%%%%%%%%%%%%%%%%%%%%%%%%%%%%%%%%%%%%%%%%%%%%%%%%%%%%%%
\subsection*{3\ Three Dimensional $BF$ Topological Theory}

Before continuing with gravity, let us recall some facts about
topological theories, specializing on the 3-dimensional $BF$ model.
The fields are a 1-form gauge potential $A$ and a 1-form field
$B$ (in matrix notation):
\eq\ba{l}
A = A_\m(x) dx^\m =  A^a_\m(x) T_a dx^\m\ ,\es
B = B_\m(x) dx^\m =  B^a_\m(x) T_a dx^\m\ ,
\ea\eqn{A,B}
where the gauge group generators $T_a$ 
obey the algebra and the trace convention
\eq\ba{l}
[ T_a,\,T_b ] =i f_{abc}T_c\ ,\quad 
Tr(T_a T_b) = \d_{ab}\ .
\ea\eqn{alg}
Introducing the ghost fields $c$, $b$, as well as the antighost 
$\bar c$, $\bar b$ and the Lagrange multiplyers fileds $\pi$, $\la$, we
may write the BRST transformations of the theory as
\eq\ba{ll}
sA_\m = D_\m c \equiv \pa_\m c + [c,A_\m]\ ,\quad 
                  &sc = \half \{c,c\} = c^2\ ,\es
sB_\m = [c,B_\m] + D_\m b\ ,\quad 
                  &sb =  \{c,b\} \ ,\es
s\bar c=\pi\ ,  \quad &s\pi=0\ ,\es
s\bar b=\la\ ,   \quad &s\la=0\ .
\ea\eqn{BRST}
The ghost and antighost fields are anticommuting scalar fields, 
whereas the Lagrange multipliers are commuting. All fields belong to 
the adjoint representation of the gauge group:
\eq
\vf(x) \equiv \vf^a(x) T_a
\eqn{adjoint}
We have directly written the gauge transformations of the theory 
in the form of BRST transformations.
The usual gauge transformations of $A$ and $B$ are the 
transformations 
\equ{BRST} with the ghost fields $b$ and $c$ replaced by ordinary local
infinitesimal c-number parameters. The transformation corresponding to
the ghost $b$ is specific of the $BF$ models.
The BRST transformations \equ{BRST} are 
nilpotent:
\[
s^2=0\ .
\]
The BRST-invariant action reads
\eq
S = S_{\rm inv} + S_{\rm gf}\ ,
\eqn{action}
where the first term is the gauge invariant action 
of the $BF$ theory~\cite{schwarz}:
\eq
S_{\rm inv} = \tr \int_{R^3} BF = 
               \half \tr\dint d^3x\, \e^{\m\n\r} B_\m F_{\n\r} \ ,
\eqn{g-inv-action}
with
\[
F_{\m\n} = \pa_\m A_\n - \pa_\n A_\m -[A_\m,A_\n] \ ,
\]
and the second term of \equ{action} is the gauge fixing action
\eq\ba{l}
S_{\rm gf} = 
- s\, \tr\dint d^3x \sqrt{g}g^{\m\n} \lp  \pa_\m \bar c\,A_\n +
 \pa^\m \bar b\,B_\m \rp \es
\phantom{S_{\rm gf}} = 
\tr \dint d^3x \sqrt{g}g^{\m\n}\lp -\pa^\m \pi\, A_\n 
   - \pa^\m \la\, B_\n  \ 
  + \pa^\m \bar c\, s A_\n + \pa^\m \bar b\, s B_\n \rp \ ,
\ea\eqn{g-fix-action}
implementing the Landau gauge conditions $\nabla^\m A_\m$ = 0 and 
$\nabla^\m B_\m$ = 0, with $\na$ the covariant derivative 
with respect to a background metric $g_{\m\n}$.

A characteristic feature of topological theories such as the
present model, in a Landau type gauge, is the invariance of the action
under {\it vector
supersymmetry}, a supersymmetry whose generators are components of a
vector. The infinitesimal vector supersymmetry transformations read -- we
set here the background metric to be flat:
\eq
\d_{\rm vector\ susy} \vf = \e^\m \d_\m \vf\ ,
\eqn{def-xi}
where $\e^\m$ are three infinitesimal Grassmann parameters, and:
\eq\ba{ll}
\d_\m c = A_\m\ , &\d_\m b = B_\m\ , \es
\d_\m A_\n = \e_{\m\n\r} \pa^\r \bar b\ ,\qquad
   &\d_\m B_\n = \e_{\m\n\r} \pa^\r \bar c\ ,\es
\d_\m \pi = \pa_\m \bar c\ , &\d_\m \la = \pa_\m \bar b\ , \es
\d_\m \bar c = 0 \ , &\d_\m \bar b = 0 \ .
\ea\eqn{v-susy}
One notes that these transformations lower the ghost number by one
unit. Together with the BRST operator, they obey the Wess-Zumino-like
superalgebra~\cite{vector-susy}
\eq
\{s,s\} = 0\ ,\quad \{\d_\m,\d_\n\} = 0\ ,\quad
\{s,\d_\m\} = \pa_\m\ .
\eqn{bf-superalg} 
Another typical feature -- in fact shared by all gauge theories
in Landau-type gauges -- are the {\it ghost 
equations}~\cite{ghost-eq}
\eq\ba{l}
\dint d^3x \lp \dfud{S}{c} - \lc\bar c,\dfud{S}{\pi}\rc 
 - \lc\bar b,\dfud{S}{\la}\rc \rp = 0 \ , 
\qquad
\dint d^3x \lp \dfud{S}{b} - \lc\bar b,\dfud{S}{\pi}\rc \rp = 0\ ,
\ea\eqn{bf-gh-eq}
which characterize the coupling of the ghost fields. 
Together with vector supersymetry, these equations imply the ultraviolet
finiteness -- or scale invariance -- of the topological models where
they hold (see~\cite{alg-ren} and references therein).

%%%%%%%%%%%%%%%%%%%%%%%%%%%%%%%%%%%%%%%%%%%%%%
\subsection*{4\ 4-Dimensional Gravity}

Whereas gravity in 3-dimensional space-time is equivalent to a topological
$BF$ theory, the gauge invariance being local Lorentz invariance, this
is no more the case in 4 dimensions and higher. However, as announced in
the Introduction, some features of topological theories are kept, such as
the ghost equations and vector supersymmetry. Let us first write the
BRST transformations leaving the Palatini action \equ{inv-action} 
invariant~\cite{BRS-grav}:
\eq\ba{l}
s E^I = \LL_\xi E^I + \om^I{}_J E^J\ ,\es
s A^{IJ} = \LL_\xi A^{IJ}
  + d\om^{IJ} + \om^I{}_K A^{KJ} + \om^J{}_K A^{IK}\ ,\es
s \xi = \half \{\xi,\xi\} \ ,\quad  
(\, \mbox{or: }s\xi^\m = \xi^\la\pa_\la \xi^\m\,) \ ,\es
s \om^I{}_J = \LL_\xi \om^I{}_J + \om^I{}_K\om^K{}_J\ ,
\ea\eqn{BRS}
with the nilpotency property $s^2=0$. 
They correspond to local Lorentz invariance and diffeomorphism
invariance. 
The infinitesimal parameters $\xi^\m(x)$ - the components of
the vector $\xi$ -- and $\om^I{}_J(x)$ 
of the respective symmetries are now Grassmann  
(i.e. anticommuting) number fields -- the Faddeev-Popov ghosts.
The bracket $\{\ ,\ \}$ is the Lie 
bracket 
$\{u,v\}^\m = u^\la\pa_\la v^\m \pm v^\la\pa_\la u^\m$,
with the sign $+$ if both $u$ e $v$ are odd, and the sign $-$ otherwise.

In order to gauge fix the theory with respect to its local symmetries --
diffeomorphism and local Lorentz invariances -- we introduce 
the respective antighost
$\bxi_I$, $\bom_{IJ}$ and Lagrange multipliers $\la_I$, $b_{IJ}$, with
the following nilpotent BRST transformations:
\eq
s\BC_i =\Pi_i\ ,\quad s \Pi_i=0\ ,\quad i =1,2\ ,
\eqn{BRS-fix}
where we are using the condensed notation
\eq\ba{ll}
\{\BC_i,\,i =1,2\} = \{\bxi_I,\,\bom_{IJ}\}    \ ,\quad  
       &\{\Pi_i,\,i =1,2\}= \{\la_I,\,b_{IJ}\}\ ,\es
\{\AA^i_\m,\,i =1,2\} = \{E_\m^I,\,A_\m^{IJ}\}\ ,\quad &
\ea\eqn{notation}
The gauge fixing part of the action is then defined as:
\eq\ba{l}
S_{gf} = -s\dintm d^4x \sqrt{-g} g^{\m\n} \dsum{i=1}{2}
\pa_\m\BC_i \AA^i_\n  \es
\phantom{S_{gf} } = \dintm d^4x \sqrt{-g} g^{\m\n} \dsum{i=1}{2} \lp 
- \pa_\m\Pi_i \AA^i_\n  + \pa_\m\BC_i s\AA^i_\n  \rp\ ,
\ea\eqn{gf-action}
which is automatically BRST invariant. Note that in order to contract the
world indices $\m$, $\n$ we had to introduce a background metric
$g_{\m\n}$ -- not to be confounded with the physical, 
dynamical metric $G_{\m\n}$ defined in \equ{metric}. 

It is worthwhile to remark that the gauge fixing action is completely
determined by its BRST invariance and the ``gauge condition''
\eq 
\dfud{S}{\Pi_i} = \pa_\m\lp \sqg g^{\m\n}\AA_\n^i \rp\ ,
\eqn{gauge-cond}
where $S$ is the total action
\eq
S = S_{\rm inv} + S_{\rm gf}\ .
\eqn{g-tot-action}
As in the $BF$ model of last section, the ghost $\xi$ and $\om$ obey
ghost equations. Let us concentrate on the equation for $\xi$:
\eq
\dfud{S}{\xi^\m} = \dsum{i=1,2}{} \lp
-\sqg g^{\n\la} \pa_\la\BC_i\pa_\m\AA^i_\n  +
\pa_\n \lp \sqg g^{\n\la} \pa_\la\BC_i \AA^i_\m\rp \rp \ .
\eqn{xi-eq}
What is interesting with this ghost equation, 
is that it may be interpreted as a Ward
identity for a {\it vector supersymmetry}, at least for some class 
of background metrics. Let us first consider the simpler case of a flat
background metric and write the space-time integral of the ghost
equation, having used integrations by parts and the gauge
condition \equ{gauge-cond}:
\eq
\d_\m S \equiv \dintm d^4x \lp \dfud{}{\xi^\m} 
+ \dsum{i=1,2}{} \pa_\m \BC_i\dfud{}{\Pi_i} \rp S = 0\ .
\eqn{susy-w-i}
Indeed, the last equation expresses
the invariance of the action under the infinitesimal transformations
\eq\ba{l}
\d_\m \xi^\n = \d_\m^\n\ ,\es
\d_\m \Pi_i = \pa_\m \BC_i\ , \es
\d_\m \vf = 0\ , \quad \vf \not= \xi^\m\,,\ \Pi_i  \ ,
\ea\eqn{vec-susy-flat}
The supersymmetry operators $\d_\m$, together 
with the BRST operator $s$, obey the superalgebra
\eq
\{s,s\}=0\ ,\quad \{ \d_\m ,\d_\n \}=0\ ,\quad \{s,\d_\m \} = \pa_\m \ ,
\eqn{susy-alg-flat}
where the partial derivative $\pa_\m$ is
the infinitesimal generator of the translations. 

In the case of a general non-flat background metric, we define 
the vector supersymmetry transformations
\eq
\dse = \e^\m \d_\m    \ ,
\eqn{vec-susy}
with $\d_\m$ given by \equ{vec-susy-flat}, and 
where the infinitesimal parameter 
$\e^\m$ is a vector field -- taken as commuting, to the 
contrary of $\xi^\m$.
The supersymmetry operators $\dse$, together 
with the BRST operator $s$, obey now the superalgebra
\eq
\{s,s\}=0\ ,\quad \{ \dse ,\d^{\,{\rm S}}_{(\e')}
\}=0\ ,\quad \{s,\dse \} = \pa_\m \ ,
\eqn{susy-alg}
where the Lie derivative $\LL_\e$ is
the infinitesimal generator of the diffeomorphisms along the 
vector field $\e$.

It turns out~\cite{piguet-cqg} that the action \equ{g-tot-action} 
is invariant
under the vector supersymmetry \equ{vec-susy} provided the infinitesimal 
parameter $\e$ is a Killing vector
of the background metric, i.e. satisfies to the condition
\eq
\LL_\e g^{\m\n}= 0\ ,\quad\mbox{or:}\quad
   \nablag_\m\e_\n + \nablag_\n\e_\m = 0\ .
\eqn{killing}
Thus, vector supersymmetry holds if the background metric admits
Killing vectors. The number of independent Killing vectors is maximum
for a flat background metric. In this case, the general solution 
of the condition \equ{killing} reads
\eq
\e^\m = a^\m + b^{\m\n}x_\n\ ,\quad \mbox{with}\quad
a^\m\,,\  b^{\m\n}=-b^{\n\m}\quad \mbox{constant parameters}\ .
\eqn{susy-poinc}
The right hand side of the anticommutator in \equ{susy-alg} is then an
infinitesimal rigid Poincar\'e transformation of parameters 
$a^\m$ and $b^{\m\n}$. This last result is in contrast to the one
obtained in the case of the topological models (see e.g.~\cite{cs-lucch-pig} 
for the 3-dimensinal Chern-Simons theory), where namely only the
vector supersymmetry associated to translation invariance is obtained in
the flat limit.

Let me finally mention that, as in the Yang-Mills case, the vector symmetry 
operator may be expressed in the form of 
the so-called operator $\d$ of 
Sorella~\cite{sorella-delta} used to construct the invariants 
of the theory, and characterized by the 
algebraic relation
\eq
[\d,s] = d\ ,
\eqn{delta-s-com}
where $d$ is the exterior derivative. A construction of the operator
$\d$ is given in~\cite{brs-grav-wien}. More details may be found 
in~\cite{piguet-cqg}.

%%%%%%%%%%%%%%%%%%%%%%%%%%%%%%%%%%%%%%%%
% BLANK LINE: 

\noindent{\bf Acknowledgements.} I am very grateful to the
members of the I.E. Tamm Theory Department of the Lebedev Physical
Institute, and specially to the organisers of this conference, for their
invitation, their very warm hospitality and their help in all kind of
practical matters. I thank very much Clisthenis Constandinis and 
Fran\c cois Gieres for interesting and useful discussions. 
The author acknowledges the Conselho Nacional de
Desenvolvimento Cient\'\i fico e 
Tecnol\'ogico (CNPq -- Brazil) for a travel grant.

%%%%%%%%%%%%%%%%%%%%%%%%%%%%%%%%%%%%%%%%%

\end{document}